\documentclass[amsmath,amssymb,aps,eqsecnum]{revtex4}
\usepackage{graphicx}
\usepackage{amsfonts}
\usepackage{bm,bbm}

\begin{document}

\title{Average fidelity between random quantum states}

\author{Karol
{\.Z}yczkowski$^{1,2,3}$ and Hans-J{\"u}rgen Sommers$^4$}
\affiliation {$^1$Instytut Fizyki im. Smoluchowskiego,
Uniwersytet Jagiello{\'n}ski,
ul. Reymonta 4, 30-059 Krak{\'o}w, Poland}
\affiliation{$^2$Centrum Fizyki Teoretycznej,
Polska Akademia Nauk, Al. Lotnik{\'o}w 32/44, 02-668 Warszawa, Poland}
\affiliation{$^3$Perimeter Institute,
Waterloo, Ontario N2L 2Y5, Canada}
\affiliation{$^4$Fachbereich Physik,
Universit\"{a}t Duisburg-Essen, Campus Essen, 
  45117 Essen, Germany}

 \date{July 16, 2004}

\begin{abstract}
We analyze mean fidelity between random density matrices
of size $N$, generated with respect to various
probability measures in the space of mixed quantum states:
Hilbert-Schmidt measure, Bures (statistical) measure,
the measures induced by
partial trace and the natural measure on the space of pure states.
In certain cases explicit 
probability distributions for fidelity are derived.
Results obtained may be used to gauge the
quality of quantum information processing schemes.
\end{abstract}

 \pacs{03.65.Ta}

\maketitle

\medskip
\begin{center}
{\small e-mail: karol@cft.edu.pl \ \quad \
sommers@theo-phys.uni-essen.de}
\end{center}

\section{Introduction}

Modern applications of quantum theory renewed
interest in characterization of the set of mixed quantum states.
It is often necessary to quantify, how a certain mixed
state may be approximated by another one.
For this purpose one may use various distances
defined in the space of mixed quantum states, e.g.
the trace distance, the Hilbert-Schmidt (HS) distance and the Bures
distance \cite{Bu69}.
The latter distance, is a function of {\sl fidelity} \cite{Jo94},
and may be considered as a generalization of the overlap between
pure states \cite{Uh76}. Fidelity between a pair of two mixed states
is equal to unity if and only if they do coincide.
Optimal fidelity between the input and the output states
are used to characterize the
 quality of a cloning algorithm \cite{BH96,Ce00}.

In various quantum problems one computes the mean fidelity,
averaging over a certain ensemble of states.
Analyzing dynamics of quantum chaotic systems one
studies the fidelity between a given initial state and its image
under the action of the quantum map \cite{Pr02}. To extract
global information on the entire system one averages
this quantity over the natural, Fubini--Study (FS) measure on the space
of initial pure states. Mean fidelity between
two images of a random pure state
obtained by the action of a certain unitary operation and a certain
irreversible, contracting superoperator $\cal E$
was found  by Bowdery et al. \cite{BOSJ01} for qubits
and by Nielsen \cite{Ni02} for the more general case of
$N$ dimensional states. Another expression  
for the average gate fidelity was recently 
derived by Bagan et al. \cite{BBMT03}, in which the nontrivial term 
was shown to be proportional to the mean value of the Hilbert--Schmidt scalar
product $\langle {\rm Tr}\gamma_i{\cal E}(\gamma_i)\rangle_i$
 of the $SU(N)$ generators $\gamma_i$ and their images
${\cal E}(\gamma_i)$, averaged over all of them, $i=1,\dots,N^2-1$.    

In this work we study a complementary problem
and analyze mean fidelity $\langle F\rangle$ between two independent,
random states.
Such a problem may be interesting 
from a practical point of view: 
The average fidelity may serve as
reference value by analyzing the  mean (maximal/minimal) fidelity 
achieved in a certain protocol of quantum cloning. 
For instance, if the measured (computed) fidelity between a certain pair
of one--qubit mixed states only slightly exceeds the average value,
one should not conclude that these states
are more correlated than two generic, random mixed states.

In this way we are in position to propose a general tool
measuring the quality of a given theoretical scheme 
of quantum information processing 
or its experimental realization.  
Let $\tilde F$ denote the mean fidelity between the state 
obtained in an analyzed cloning protocol and the target state.
Then the quality of the protocol may be gauged 
by a dimensionless coefficient 
$\alpha=[{\tilde F}-\langle F\rangle]/
\sqrt{\langle F^2\rangle - \langle F\rangle^2}$, 
where $\langle F\rangle$ denotes the average over
an appropriate ensemble of random density matrices.

We are also interested in the
probability distribution $P(F)$. Two averaging schemes should be
distinguished. In the symmetric case, both states are drawn at random
according to the same measure
(e.g. both pure states or both mixed states generated with respect to the same
probability distribution). In the non-symmetric case,
both probability distributions are different:
in particular we study the mean fidelity between a random pure state
and a random mixed state, generated with respect to a
specified measure.

Let us emphasize in this place that there exists
no single, naturally distinguished probability measure
in the set of mixed quantum states \cite{Ha98,ZS01,SZ03b}.
Guessing a mixed state of size $N$ at random we might
use additional information, if available.
For instance, knowing that the mixed
state has arisen by the partial tracing
over a $K$ dimensional environment, 
the induced measure \cite{ZS01} should be used.
In particular, if the size of the environment $K$
and the size $N$ of the system are equal, then 
the measure induced by partial trace 
coincides with the Hilbert-Schmidt measure.
Without any information concerning the random
state whatsoever, it will be legitimate
to make use of the Bures measure,
related to the Jeffrey's prior, statistical distance 
and the distinguishability.

 The paper is organized as follows.
In section 2 we review the basic properties of fidelity, while
in section 3 we introduce the necessary measures in the set
of mixed quantum states of size $N$.
Mean fidelity between pure and mixed states are
analyzed in section 4. Main 
results of this work are presented in section 5, in which
we  compute the mean fidelity, averaged over two independent
random mixed states, generated with respect to 
an arbitrary induced measure,
 labeled by the size $K$ of the environment.
The probability distribution for fidelity between
a random pure state and a random mixed state distributed
according to the Bures measure is computed in Appendix B, while
the derivation of the generating functions for the moments
of the root fidelity for two mixed states of arbitrary $N$
generated independently according to 
an induced measure is postponed to Appendix A.

\section{Fidelity}

Let ${\cal M}_N$ denote the set of all mixed quantum states
of size $N$. It contains all Hermitian, semipositive definite
matrices of size $N$,
which are trace normalized,
\begin{equation}
{\rm Tr} \rho=1 \ .
\label{trace}
\end{equation}
We are going to consider two distances in this set:
the {\sl Hilbert-Schmidt distance}
\begin{equation}
D_{HS}(\rho_1,\rho_2)=\sqrt{ {\rm Tr}   [(\rho_1 - \rho_2)^2] }.
  \label{HilbSchmi}
\end{equation}
and the {\sl Bures distance} \cite{Bu69,Uh76},
 \begin{equation}
 \bigl( D_{B}(\rho_1,\rho_2)\bigr)^2 = 2 - 2\
\mbox{Tr}\sqrt{\sqrt{{\rho}_1}
{\rho}_2\sqrt{{\rho}_1} }
 = 2 - 2\sqrt{F(\rho_1,\rho_2)} \ .
 \label{Bures}
\end{equation}
respectively.

The Bures distance is distinguished by several remarkable properties:
for pure states it agrees with the natural, Fubini-Study distance,
while in the space of diagonal density matrices it induces
the statistical (Fisher) distance   \cite{PS96}.
The Bures distance is a function of {\sl fidelity} \cite{Jo94}:
\footnote{Although this was the original
definition of Jozsa, some authors use this name for $\sqrt{F}$.}.
\begin{equation}
 F(\rho_1,\rho_2) :=
\Bigl[ \mbox{Tr}\sqrt{\sqrt{{\rho}_1}
{\rho}_2\sqrt{{\rho}_1} } \Bigr]^2 .
 \label{fidel}
\end{equation}
This quantity is sometimes called
{\sl Uhlmann transition probability} \cite{Uh76}, since
 for a pair of pure states it reduces to the squared
overlap,
 $F={\rm Tr}\rho_1\rho_2=|\langle \psi_1|\psi_2 \rangle |^2$

Fidelity enjoys several important properties \cite{Jo94}:
it is a symmetric,
non-negative, continuous, concave function of both states,
unitarily invariant,
equal to unity if and only if both states do coincide.
Therefore it becomes an important tool
to characterize the closeness of any two mixed states,
often used in modern applications of quantum mechanics
(see e.g. \cite{NC00}).
The only disadvantage consists in computing it explicitly:
to find the fidelity one needs to diagonalize the density matrices,
 but more importantly,
 the fidelity stays being a function of the eigenvectors.

Apart of the fidelity $F$, we will also use the square root
fidelity $\sqrt{F}$. Also this quantity satisfies 
several appealing properties \cite{NC00} and
in certain cases its probability distribution is
easier to compute then $P(F)$. Furthermore the quantity $\arccos[\sqrt{F(\sigma,\rho)}]$
 has a clear geometric interpretation
as {\sl Bures angle} in the space of mixed states 
equipped with the Bures metric. 
It is equal to the geodesic Riemannian distance
between both mixed states \cite{Uh92,Uh95},
and in the case of pure states
it coincides with the Fubini--Study distance.

Computation of fidelity is much simpler for $N=2$.
To derive an explicit formula for fidelity 
we use first steps of the calculation of the Bures
distance presented by  H{\"u}bner \cite{Hu92}.
Any two-by-two matrix $A$ obeys the characteristic equation
\begin{equation}
 A^2 - A\mbox{Tr}A +{\mathbbm 1}  \det{A} = 0 \ .
\end{equation}
Taking the trace we obtain
\begin{equation}
(\mbox{Tr}A)^2 = \mbox{Tr}A^2 + 2\det{A} \ .
\label{trac2}
\end{equation}
Let us now set $A= \sqrt{\sqrt{{\rho}_1}{\rho}_2 \sqrt{{\rho}_1}}$,
so that  Tr$A^2={\rm Tr}{\rho}_1{\rho}_2$ and
$\det A=\sqrt{\det{{\rho}_1}\det{{\rho}_2}}$.
Let $(p_1,p_2)$ denote the spectrum of a state $\rho$.
Hence $1=(p_1+p_2)^2={\rm Tr}\rho^2+2 \det \rho$
and $2\det \rho=1-{\rm Tr}\rho^2$. Substituting this into
Eq. (\ref{trac2}) we obtain
\begin{equation}
 F(\rho_1,\rho_2)= ({\rm Tr}A)^2 =
 \mbox{Tr} \rho_1 \rho_2 +
 \sqrt{ (1- \mbox{Tr} \rho_1^2 ) (1- \mbox{Tr} \rho_2^2 )} \ .
 \label{FidN2}
\end{equation}

\section{Measures in the space of mixed quantum states}

Random quantum states may be generated according to
different measures, and it is
hardly possible to distinguish the unique
probability measure in the set of density matrices.
Usually one considers product measures,
which may be factorized \cite{Ha98,ZS01}
\begin{equation}
{\rm d} \mu_x = {\rm d} \nu_x
 (\lambda_1,\lambda_2,...,\lambda_N) \times {\rm d} h \ .
  \label{product}
\end{equation}

The second factor d$h$, determining the distribution of the eigenvectors of
the density matrix, is the unique, unitarily invariant Haar measure on
$U(N)$,  while the first factor  depends only on the
eigenvalues $\lambda_i$ of $\rho$ and may be chosen by an arbitrary
probability distribution d$\nu_x=P_x({\vec \lambda}) d^N \lambda$.
The joint probability distribution
is defined on the $(N-1)$ dimensional simplex $\Delta_{N-1}$,
which contains all vectors of non-negative
entries summing to unity.

The Hilbert-Schmidt measure, induced by the
Hilbert-Schmidt metric, is defined by the
following joint probability distribution \cite{Ha98,ZS01}
\begin{equation}
 P_{\rm HS}^{(N)} ({\vec \lambda}) =
 \frac{\Gamma(N^2)\  \delta\bigl(\sum_{j=1}^N \lambda_j -1 \bigr) }
{\prod_{j=0}^{N-1} \Gamma(N-j) \Gamma(N-j+1) }
\prod_{i<j}^N (\lambda_i-\lambda_j)^2 .
  \label{HSmes}
\end{equation}

The Bures metric leads to the Bures
measure in the simplex of eigenvalues
\begin{equation}
 P_{\rm B}^{(N)}({\vec \lambda}) =
\frac{   2^{N^2-N}\  \Gamma(N^2/2)}{\pi^{N/2}\ \Gamma(1)...\Gamma(N+1)}
\frac{\delta\bigl( \sum_{j=1}^N  \lambda_j -  1\bigr)}
       {\sqrt{\lambda_1\lambda_2 \cdots \lambda_N}}
   \prod_{i<j}
\frac{ (\lambda_i-\lambda_j)^2 }
     {\lambda_i+\lambda_j} .
  \label{mesbur2}
\end{equation}
This  probability distribution was derived by Hall \cite{Ha98},
while the normalization constants where found in \cite{Sl99b}
for $N=3,4,5$ and in \cite{SZ03} for an arbitrary $N$.

Is is also instructive to consider a family
of measures in the space
of mixed states of size $N$ induced by the Haar measure
on the unitary group $U(NK)$. The integer parameter $K$,
used to label the measure $\mu_{N,K}$, represents the size of
an ancilla.  A mixed state of size $N$ may be obtained by tracing a 
certain random pure state $|\Phi\rangle$
of size $NK$ over the $K$--dimensional ancilla,
$\rho={\rm T}_K\bigl(|\Phi\rangle \langle \Phi|\bigr)$.
Representing the pure state $|\Phi\rangle$
in an arbitrary product basis 
$|i,j\rangle=|i\rangle \otimes |j\rangle$
where $i=1,...,N$ and $j=1,...,K$
we obtain $|\Phi\rangle = \Phi_{ij} |i,j\rangle$.
The rectangular  matrix of coefficients $\Phi_{ij}$
of size $N \times K$ allows us to write
the reduced state as $\rho=\Phi \Phi^{\dagger}$. 
If matrices $\Phi$ are random, the matrices 
$\rho$ constructed in such a way are
called {\sl Wishart matrices}.
The probability distribution of eigenvalues
of $\rho$ reads \cite{Lu78,LS88,Pa93,ZS01}
\begin{equation}
 P_{\mu_{N,K}}({\vec \lambda}) =
\frac{\Gamma(KN)
\delta\bigl( \sum_{j=1}^N  \lambda_j -  1\bigr)}
{\prod_{j=0}^{N-1} \Gamma(K-j) \Gamma(N-j+1) }
\prod_i\lambda_i^{K-N}
\prod_{i<j}(\lambda_i-\lambda_j)^2
  \label{HSindNK}
\end{equation}
and for $K=N$ reduces to the Hilbert-Schmidt  measure (\ref{HSmes}).
In the above equation the inequality $K\ge N$
is assumed and this case is called the Wishart case.
If this is not the case, (the so--called anti--Wishart case 
\cite{YZ02,JN03}) 
induced measures with $K<N$ are singular, since
they are supported by the subspace of states of submaximal rank $K$
which belongs to the boundary of ${\cal M}_N$.
In particular, the measure $\mu_{N,1}$
is just the natural Fubini--Study  measure
on the space of pure states in ${\cal H}_N$,
since the partial trace over an $1$--dimensional ancilla
does not change the pure state.

Interestingly, in the one--qubit case,
for $K=3/2$, the induced measure
  (\ref{HSindNK}) reduces
to the Bures measure  (\ref{mesbur2}),
\begin{equation}
 P_{\rm B}^{(2)}({\vec \lambda}) =
P_{\mu_{2,3/2}}({\vec \lambda}) 
  \label{indbur32}
\end{equation}
Although such a coincidence occurs for an
non-integer value
of the dimensionality  of the environment,
and seems not to have any physical meaning,
it will help us to obtain some results
for the Bures measure
by analytical continuation in $K$.
Unfortunately this
trick works only in the  $N=2$ case,
for which $\lambda_1+\lambda_2=1$,
so the denominator under the product in
the distribution (\ref{mesbur2}) is
trivially equal to unity.

We are going to study mean fidelity $\langle F\rangle$
with respect to different measures.
Let $\langle F\rangle_{\mu}$ denote the homogeneous case, in which
both arguments of fidelity are random mixed states, distributed independently
with respect to the same probability measure $\mu$.
In a more general case we may average over
two different measures and such averages will be denoted by
$\langle F\rangle_{\mu_1}^{\mu_2}$.
Since fidelity is symmetric
$F(\rho_1,\rho_2)=F(\rho_2,\rho_1)$,
we have $\langle F\rangle_{\mu_1}^{\mu_2}=
\langle F\rangle_{\mu_2}^{\mu_1}$.

An analogous notation will be also
used to label probability distributions for fidelity:
$P_{\mu}(F)$ denotes the distribution for the symmetric case,
in contrast to $P_{\mu_1}^{\mu_2}(F)$,
in which both states are averaged over different measures.
To simplify notation, instead of
writing $\mu_{N,K}$ as a label denoting a certain induced measure,
we shall use the  labels $N,K$.

\section{Mean fidelity: non--symmetric averaging}

\subsection{One state pure, one arbitrary}

Computation of fidelity between two states simplifies considerably,
if one of the states is pure, $\rho_1=|\psi_1\rangle \langle \psi_1|$,
\begin{equation}
F\bigl( |\psi_1\rangle \langle \psi_1|,\rho_2\bigr)=
{\rm Tr} |\psi_1\rangle \langle \psi_1|\rho_2 \ .
\label{fidpure}
\end{equation}
Working in the basis which contains $|\psi_1\rangle$
we see that fidelity is equal to the matrix
element $\rho_{11}:=\langle \psi_1|\rho_2 |\psi_1\rangle$.
We are going to analyze the case in which the
random pure state $|\psi_1\rangle$
is  generated according to the natural,
Fubini--Study measure on the space of pure states.
For any mixed state represented in a random basis all components have, on
average,
the same magnitude, hence
 \begin{equation}
\langle F\rangle_{\rm FS}= \langle F\rangle_{N,1}=
\frac{1}{N},
\label{pureN}
\end{equation}

If the second random state is also pure,
$\rho_2=|\phi\rangle \langle \phi|$,
then the fidelity is equal to the overlap between both states,
$F=|\langle \psi|\phi\rangle|^2$.
Assuming both states to be generated independently
according to the same FS measure we infer that
fidelity is equal to the squared component of a
random vector in a random basis.
The probability distribution for this quantity
 \begin{equation}
 P_{N,1}(F) = (N-1) (1-F)^{N-2},
\label{ditspureN}
\end{equation}
was derived in \cite{KMH88} while analyzing
eigenvectors of a random unitary matrix
pertaining to circular unitary ensemble (CUE)
and distributed according to the Haar measure on $U(N)$.
Note that this result holds also in the case if
one of the pure states (say the second one) is fixed,
$P(\rho_2)=\delta(\rho_2-|\psi_*\rangle \langle \psi_*|)$.

Let us now consider a non-symmetric averaging:
one state $\rho_1=|\psi_1\rangle \langle \psi_1|$ is pure and is
distributed according to the FS measure, while the other one, $\rho_2$, is
distributed according to
the measure $\mu_{N,K}$ in the space of mixed states. Hence the latter state
may be obtained by tracing a certain random pure state $|\Phi\rangle$
of size $KN$ over the $K$--dimensional ancilla.
Fidelity between them, $F\bigl(|\psi\rangle \langle \psi|, \rho_2 \bigr)$,
is given by the sum of $K$ terms,
$F=\rho_{11}=\sum_{i=1}^K |c_i|^2$
where $c_i$, $i=1,...,KN$ denote the components of the
pure state $|\Phi\rangle$.
In our previous work we have analyzed $M$--dimensional
truncations of complex random vectors of dimensionality $L$ and
unit length. The length $t$ of the truncated vector was
shown to be distributed according to $P(t)\sim t^{2M-1} (1-t^2)^{L-M-1}$
 \cite{ZS00}. In the case considered the fidelity is just equal to
the squared length, $F=t^2$, the initial length of the vector $L=KN$,
and the length of the truncation $M$ is equal to $K$. Hence
changing variables and finding
the normalization constants in terms of the Euler Gamma function we
arrive at the probability distribution
\begin{equation}
 P_{N,K}^{N,1}(F) = \frac{\Gamma(KN)}
                          {\Gamma(K) \Gamma[K(N-1)]}
 F^{K-1} (1-F)^{K(N-1)-1}
\label{ditsmixN}
\end{equation}
describing the fidelity $F$ between a random pure state of size $N$
and a random mixed state generated according to the measure $\mu_{N,K}$.
In the special case $K=1$ the second state is also pure, and
the above formula reduces to (\ref{ditspureN}). In the case
 $K=N$  the mixed state is distributed according to
the Hilbert-Schmidt measure, and the probability distribution reads
\begin{equation}
 P_{N, HS}(F) = \frac{\Gamma(N^2)}
                          {\Gamma(N) \Gamma[N^2-N]}
 F^{N-1} (1-F)^{N(N-1)-1}
\label{ditsmixHSN}
\end{equation}

For completeness, let us formulate an analogous result
following from studying truncations
of real random vectors \cite{ZS00},
\begin{equation}
 P_{N,K}^{N,1,\mathbb R}(F) = \frac{\Gamma(KN/2)}
                          {\Gamma(K/2) \Gamma[K(N-1)/2]}
              F^{K/2-1} (1-F)^{K(N-1)/2-1}.
 \label{ditmixNRe}
\end{equation}
This distribution characterizes
fidelity $F$ between a {\sl real} random pure state of size $N$
and a {\sl real} random mixed state generated according to the
induced measure.

Let us examine in some detail the special case
of $N=2$. If $F$ denotes the fidelity
between $\rho_2$ and $|\psi_1\rangle\langle \psi_1|$  
than the fidelity of the same mixed state $\rho_2$
with respect to any pure state orthogonal to $|\psi_1\rangle$
is equal to $1-F$.
Since we average over the entire Bloch sphere
of pure states, the probability distribution will
be a symmetric function of $F$ and $(1-F)$.
This is indeed the case and for $N=2$
the distribution (\ref{ditsmixN})
reduces to $P_{2,K}^{2,1}\sim F^{K-1}(1-F)^{K-1}$.
In particular, for the HS measure, $(K=2)$,
and the Bures measure $(K=3/2)$,  one obtains
\begin{equation}
P_{2,HS}=6 F(1-F)
{\quad \rm and \quad}
P_{2,B}=\frac{8}{\pi} \sqrt{F(1-F)},
\label{burhs2}
\end{equation}
respectively.
Eq. (\ref{ditmixNRe})
implies analogous results for the rebits,
\begin{equation}
P_{2,HS}^{\mathbb R}=1
{\quad \rm and \quad}
P_{2,B}^{\mathbb R}= 1\ .
\label{burhs2R}
\end{equation}
Note, that for rebits $P_{2,HS}^{\mathbb R}$ and $P_{2,B}^{\mathbb R}$
coincide,
 while $ K=3/2 $ has no direct physical meaning in this case.

\begin{figure} [htbp]
   \begin{center}
 \includegraphics[width=4.2cm,angle=0]{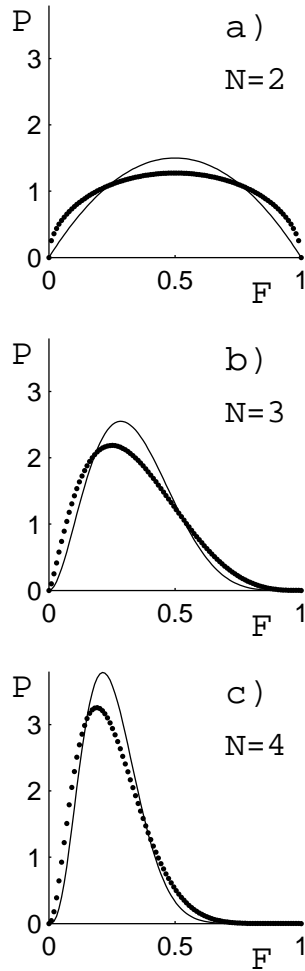}
 \vskip -0.2cm
\caption{Probability distribution of fidelity
between a random pure state and a random mixed state
distributed according to the Hilbert--Schmidt measure -
eq. (\ref{ditsmixHSN}), solid line
and Bures measure - 
eq. (\ref{ditsmixBuN}) - bold dotted line,  for 
a) $N=2$, b) $N=3$ and c) $N=4$.}
 \label{fig0}
\end{center}
 \end{figure}

In the general case of an arbitrary $N$ it is also possible 
to analyze the fidelity between a random pure state
 and a random mixed state, distributed according to the Bures measure.
The probability distribution, 
in a sense  complementary  to Eq. (\ref{ditsmixHSN}),
\begin{equation}
 P_{N, B}(F) = \frac{\Gamma(N^2/2) \Gamma(2N-1)}
                          {\Gamma(N) [\Gamma(N-1/2)]^2 \Gamma(N^2/2-N)}
 F^{N-1} \int_F^1 \frac{(x-F)^{N^2/2-N-1}(1-x)^{N-3/2} dx}{x^{N^2/2-N+1/2}}
\label{ditsmixBuN}
\end{equation}
is derived in Appendix B (to understand the method, it is more helpful 
first to read section V.D and appendix A).
In the case $N=2$ the integral diverges. However this divergence  is 
compensated by the last factor, $\Gamma(0)$, in the denominator, 
so it is not too difficult (by partial integration of the first factor under the integral)
to show that (\ref{ditsmixBuN}) reduces in this case 
to the second formula of (\ref{burhs2}).
A comparison of distributions of fidelity
between a pure state and a mixed state generated 
according to HS measures and Bures measures is presented in Fig. \ref{fig0}.
Observe that for the distributions for the Bures measure are broader since
the HS measure is more concentrated in the vicinity of maximally mixed state.

\subsection{One state maximally mixed, one arbitrary}

Let us now analyze another special case, if one state is maximally
mixed, $\rho_1=\rho_*:={\mathbbm 1}/N$. Hence
the fidelity with respect to any state $\rho_2=\rho$
reduces to
\begin{equation}
F\bigl( \rho_*, \rho \bigr) =
\frac{1}{N} \bigl( {\rm Tr} \sqrt{\rho} \bigr)^2
\label{fidmaxmix}
\end{equation}
It is then convenient to study the mean root fidelity
\begin{equation}
\sqrt{ F\bigl( \rho_*, \rho \bigr)} =
\frac{1}{\sqrt{N}}{\rm Tr} \rho^{1/2},
\label{fidmaxmi2}
\end{equation}
which may be written as a function of the generalized R{\'e}nyi entropy
of order one half.

Let us now assume that the random state $\rho$ is distributed according to
the Hilbert--Schmidt (\ref{HSmes}) or Bures (\ref{mesbur2}) measure.
Average moments for these ensembles of random density matrices
were analyzed in \cite{ZS01,SZ03b},
in which we derived asymptotic formulae
 \begin{equation}
  \langle {\rm Tr} \rho^q \rangle_{\rm HS} =
   N^{1-q} \frac{\Gamma(1+2q)}{\Gamma (1+q) \Gamma(2+q)}
  \Bigl( 1+O\bigl(\frac{1}{N}\bigr) \Bigr),
 \label{traceHS}
 \end{equation}
for the HS measure and
 \begin{equation}
  \langle {\rm Tr} \rho^q \rangle_{\rm B} =
   N^{1-q} 2^q  \frac{ \Gamma[(3q+1)/2] }
                  {\Gamma [(1+q)/2] \Gamma(2+q)}
  \Bigl( 1+O\bigl(\frac{1}{N}\bigr) \Bigr).
 \label{traceB}
 \end{equation}
for the Bures measure. Substituting $q=1/2$
we find asymptotic results for the fidelity of a random
mixed state with respect to the maximally mixed state
\begin{equation}
\Big\langle \sqrt{ F\bigl( \rho_*, \rho \bigr)} \Big\rangle_{\rm HS} =
\frac{8}{3\pi} \approx 0.843
\label{fidmaxHS}
\end{equation}
for the HS measure and
\begin{equation}
\Big\langle \sqrt{ F\bigl( \rho_*,\rho \bigr)} \Big\rangle_{\rm B} =
\frac{\sqrt{2} \Gamma(5/4)}
{\Gamma(3/4) \Gamma(5/2)} \approx 0.787
\label{fidmaxBur}
\end{equation}
for the Bures measure.
Observe that both results, valid in the limit $N \to \infty$,
converge asymptotically
to some nontrivial constants, independent of $N$.
Comparison of the Bures and the Hilbert-Schmidt measures
reveals that the former favors states of a larger purity \cite{ZS01},
hence the mean value with respect to Bures measure (\ref{fidmaxBur})
 is smaller then the analogous result (\ref{fidmaxHS})
for the HS measure.

\section{Mean fidelity: symmetric averaging}

We are going to analyze a symmetric problem
of computing average fidelity between two random states,
{\sl both} of which are generated according to the same
probability distribution covering the entire set of mixed states.

\subsection {Case $N=2$, average values}

 For $N=2$ the
problem simplifies considerably since an explicit formula (\ref{FidN2})
may be used. The expectation value
 $\langle {\rm Tr} \rho_1\rho_2\rangle=1/2$
if $\rho_1$ and $\rho_2$ are independent random mixed states
generated with respect to any product measure (\ref{product}).
To show this let us write both states in the
Pauli matrix representation
\begin{equation}
\rho =  \frac {\mathbbm 1}{2} +
{\vec \tau}\cdot {\vec \lambda} \ ,
\label{Pauli}
\end{equation}
The vector  of the normalized Pauli matrices,
${\vec \lambda}={\vec \sigma}/\sqrt{2}$,
together with the rescaled identity matrix,
${\mathbbm 1}/\sqrt{2}$,
 form an orthonormal basis
in the set of two times two complex matrices
in sense of the Hilbert-Schmidt scalar product,
$\langle A |B \rangle ={\rm Tr}A^{\dagger}B$.
Expanding in this basis the scalar product
\begin{equation}
{\rm Tr} \rho_1\rho_2 =\frac{1}{2}+{\vec \tau_1}\cdot{\vec \tau_2}
\label{trrhorho}
\end{equation}
we see that due to symmetry of the distribution of the orientation
of the Bloch vector ${\vec \tau}$
the second term does not contribute to the average
$\langle {\rm Tr}\ \rho_1\rho_2\rangle$.
For any $N=2$ mixed state one has Tr$\rho^2=1/2-r^2$
where $r=|\vec \tau|$ denotes the length of the Bloch vector
and $r\in [0,1/\sqrt{2}]$. Thus
the fidelity may be expressed by \cite{Hu92} (see eq.(\ref{FidN2}))
\begin{equation}
 F(\rho_1,\rho_2)=
 {\vec \tau_1} \cdot  {\vec \tau_2} + \frac{1}{2}
+  \sqrt{ \frac{1}{2}- r_1^2}\sqrt{ \frac{1}{2}- r_2^2} ,
\label{Fr1r2}
\end{equation}
so to compute the mean fidelity
averaged over {\sl both} random mixed states
distributed according to a given measure in the space of mixed
states it is sufficient to
find the expected value of the following function of the radius
$t=\sqrt{\frac{1}{2}- r^2}$,
 averaged with respect to an
appropriate measure. The Hilbert-Schmidt measure
covers the entire Bloch ball uniformly, so
$P_{HS}(r)=6\sqrt{2}r^2$ \cite{Ha98} and it is straightforward to
find the average $\langle t\rangle_{HS}$ which leads to
\begin{equation}
\langle F\rangle_{HS}=
\frac{1}{2}+\langle t \rangle_{HS}^2= \frac{1}{2}+\frac{9}{512}\pi ^{2} \approx
0.6735
\label{meanFHS}
\end{equation}
In the same way we may evaluate the mean fidelity
 $\langle F\rangle_{\mu_{2,K}}$ with respect to the
induced measures obtained by
the partial trace of pure states of the $2K$
dimensional Hilbert space with an arbitrary natural $K$.
The probability distribution for eigenvalues, given explicitly in \cite{ZS01},
implies the following radial distribution,
\begin{equation}
 P_K(r)=  \frac{2^{K+3/2}\Gamma(K- \frac{1}{2})}
               {\sqrt{\pi}\  \Gamma(K-1) } \
 (\frac{1}{2}-r^2)^{K-2}\  r^2 \ .
\label{PKvonr}
\end{equation}
Expectation values of
$t$ may be readily expressed by the Euler Gamma function
and allow us to compute the average fidelity.
For instance, in the case of the $K=3$
induced measure we obtain
\begin{equation}
\langle F\rangle_{2,3}=\frac{1}{2}+\left(
\frac{15}{128}\sqrt{2}\pi \right) ^{2} \approx 0.7711 ,
\label{meanK3}
\end{equation}
while the general result for any $K\ge 2$ reads
\begin{equation}
\langle F\rangle_{2,K}=\frac{1}{2}+\frac{1}{2}\left(
\frac{\Gamma(K+\frac{1}{2}) \Gamma(K-\frac{1}{2} ) }
   { \Gamma(K+1) \Gamma(K-1) } \right)^2
\label{meanKN}
\end{equation}
and for $K=2$ reduces to (\ref{meanFHS}).
In the limit of an infinitely large ancilla, $K\to \infty$,
the double quotient of the $\Gamma$ functions tends to unity,
so $\lim_{K\to \infty} \langle F\rangle _{2,K} =1$.
This result is rather intuitive, since in this limit
all random states tend to be close to the maximally mixed state $\rho_*$.

In the case of the Bures measure, the states of larger purity
(larger radius $r$) are preferred and
$P_{B}(r)=\frac{8}{\pi }
\frac{r^{2}}{\sqrt{\frac{1}{2}-r^{2}}}$
\cite{Ha98,ZS01}.
Note, that the expression (\ref{PKvonr})
includes the above case of the
Bures metric by analytic continuation: $ K \to 3/2$.
Computing the mean value $\langle t\rangle_B$
we arrive at the result
\begin{equation}
\langle F\rangle_B= \frac{1}{2}+ \langle t \rangle_{B}^2=\frac{1}{2}+
\frac{8}{9\pi ^{2}} \approx 0.590,
\label{meanFB}
\end{equation}
which was independently obtained by Bagan et al. \cite{BBMTR03}
and is a particular case of
(\ref{meanKN}) with $K=3/2$.

It is worth to emphasize that for any product measure
in the space of qubits
the average fidelity is not smaller than $1/2$.
The equality occurs for the Fubini--Study measure,
$\langle F\rangle_{FS}=1/2$,
since this measure is
concentrated exclusively on the pure states,
$P_{FS}(r)=\delta(r-R)$, so that the
last term in (\ref{Fr1r2}) vanishes. Here
$R$ stands for the Hilbert-Schmidt radius
of the Bloch Ball, $R=\sqrt{2}/2$.

\subsection {Case $N=2$, probability distribution}

In order to calculate the probability distribution $P_{2,K} (F)$
we are going to use Eq.(\ref{Fr1r2})
 and integrate out the angle $\vartheta$ between
   $\vec \tau_1$ and $\vec \tau_2$ and the radia $r_i\in [0,R]$
\begin{equation}
 P_{2,K}(F)= \frac{1}{2}\int_0^R \! \!\! dr_1 \! \int_0^R \! \! \! dr_2 \!
\int_0^{\pi}
\! \!\sin\vartheta d\vartheta P_K(r_1) P_K(r_2)
 \delta\Bigl[ F-r_1r_2\cos\vartheta-\frac{1}{2}-
\sqrt{ \frac{1}{2}- r_1^2}\sqrt{ \frac{1}{2}- r_2^2} .
 \Bigr]
 \label{P2,KvonF}
\end{equation}
The radial distribution
$ P_K(r)$
for the  measure induced by tracing over $K$--dimensional
environment is given in (\ref{PKvonr}).

  One first integrates out the angle $\vartheta$.
  Using the property of the step function
  $\theta(x-F)-\theta(y-F)= \theta((x-F)(F-y))\ \  {\rm for}\ \
  x>y$ and performing a partial integration one finds:
 \begin{equation}
 P_{2,K}(F)= C(K)
  [F(1-F)]^{2(K-1)}\
  \int_0^1  \frac{{\rm d}x}{x}\
   (x+\frac{1}{x} +2(1-2F))^{2(1-K)}
\label{P2,KvonF1}
\end{equation}
with
\begin{equation}
C(K)=2(K-1)\ \left( {\Gamma(2K)\over \Gamma(K)^2}\right)^2\ .
\label{CvonK}
\end{equation}
We see that the expression is analytic in $K$.
For $K \to 1$ we obtain $P_{2,1}(F) =1= {\rm const.}$
 The last integral can be done for an integer $2(K-1)$ and we obtain
\begin{equation}
 P_{2,K}(F)= C(K) \left({F(1-F) \over 4}\right)^{2(K-1)} {1\over
(2K-3)!}\left(d\over dF\right)^{2K-3}{\arccos(1-2F) \over \sqrt{F(1-F)}}\ .
\label{P2,KvonF2}
\end{equation}
Interestingly the most simple case is obtained for the Bures metric
\begin{equation}
 P_{2,B}(F)=
 P_{2,3/2}(F)=
   {16 \over \pi ^2} \sqrt{F(1-F}) \  \arccos(1-2F)\ ,
\label{P2,B}
\end{equation}
while for the Hilbert-Schmidt metric ($K=2$) we obtain:
\begin{equation}
 P_{2,2}(F)={9\over 2}F(1-F) -{9\over4}\sqrt{F(1-F)}
  \  (1-2F)\arccos(1-2F)\ .
\label{P2,2}
\end{equation}

\begin{figure} [htbp]
   \begin{center}
 \includegraphics[width=9.4cm,angle=0]{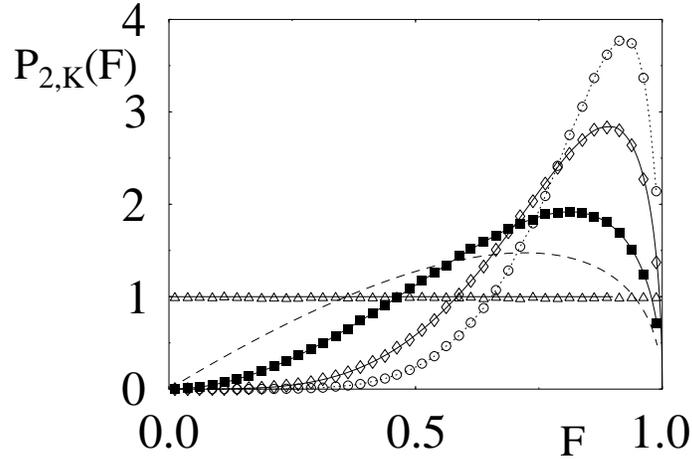}
 \vskip -0.2cm
\caption{Probability distribution of fidelity
 for random one--qubit states generated according to 
the induced measures $\mu_{2,K}$ where
$K=1$ ($\triangle$),  $K=2$ ($\square$),
and $K=3$ ($\Diamond$) $K=4$ ($\circ$).
Flat distribution represents averaging over pure states, while 
full symbols refer to the Hilbert--Schmidt measure, $K=N=2$.
Solid lines represent Eq. (\ref{ditspureN}) and 
(\ref{P2,KvonF1}), while dashed line denotes the case $K=3/2$
corresponding to the Bures distribution (\ref{P2,B}).}
 \label{fig1}
\end{center}
 \end{figure}

These probability distributions are presented in Fig. \ref{fig1}
together with exemplary numerical results. 
The behavior for $K \to \infty$ may be obtained
by a saddle-point calculation, which we start from the form
\begin{equation}
 P_{2,K}(F)= C(K) [F(1-F)]^{2(K-1)}\ {1\over2}
  \int_0^{\infty}{dx\over x}\ [x+{1\over x} +2(1-2F)]^{2(1-K)}
\label{P2,KvonF3}
\end{equation}
using the symmetry of the integral with respect
 to the inversion $x \to 1/x$.
  We find as asymptotic form for $K \to \infty$
\begin{equation}
 P_{2,K}(F)=  {\Gamma(2K+1/2)\over \Gamma(3/2)\Gamma(2K-1)}\
 F^{2(K-1)} \ \sqrt{1-F}\  \bigl[ 1+\ O\bigl(\frac{1}{K}\bigr) \bigr]\ .
\label{P2,KvonF5}
\end{equation}
As expected, for large dimensionality $K$ this distribution
tends to a delta function located
at $\langle F\rangle$,
which in the limit $K\to \infty$
tends to unity according to (\ref{meanKN}).

\subsection {General case, $N \ge 3$, average values}

In the general case  one has to use the general formula
for fidelity, (\ref{fidel}), not as convenient
for analytical computations as
Eq. (\ref{FidN2}) valid for $N=2$.
Before describing the analytical results let
us compare the problem for an arbitrary $N$
with the symmetric case,
in which the averaging in both arguments is performed
over the set of pure states.
For any nonsingular measure covering the entire set
of mixed states,
the mean value will be larger than $1/N$, since
the mean distance between random mixed states
is smaller than the mean distance between
random pure states.

This statement concerns in particular all induced measures
 $\mu_{N,K}$ analyzed for a fixed system size $N$.
 Since the mean purity decreases with the size $K$ of the ancilla,
$\langle {\rm Tr}\rho^2\rangle=(N+K)/(NK+1)$ \cite{ZS01},
one may infer that
$\langle F\rangle_{N,K} > \langle F\rangle_{N,L}$
if $K>L$. In the analogy to the simplest case $N=2$
one may thus expect that for any $N$ the
mean fidelity tends to unity,
$\lim_{K\to \infty} \langle F\rangle _{N,K} =1$.

Veracity of this reasoning
may be checked by analysis of the following
general results, derived in the
Appendix A for an arbitrary induced measure
$\mu_{N,K}$.
We are going to study the moments of the
root fidelity $\langle \sqrt{F}\rangle_{N,K}$,
so the second moment gives the average fidelity.
Let us denote $n=K-N\ge 0$ and
introduce constants
\begin{equation}
G(m) = \Bigl( \frac{\Gamma(KN)} {\Gamma(KN+m/2)} \Bigr)^2 \ .
\label{constG}
\end{equation}
Defining an auxiliary matrix of size $N$
\begin{equation}
(X_n)_{k,l}:= \Gamma(n+k+l-1) \Gamma (n+l)
{\rm \quad for \quad}
k,l = 1,2,...,N.
\label{Xnkl}
\end{equation}
 the rather complicated averages
may be written down in a concise way,
\begin{equation}
\langle \sqrt{F}\rangle_{N,K} = 
 G(1)\ {\rm Tr}[X_n^{-1} X_{n+1/2}]
\label{sqrtF}
\end{equation}
and
\begin{equation}
\langle F\rangle_{N,K} =
 G(2) \Bigl[ {\rm Tr}\bigl( X_n^{-1} X_{n+1} \bigr) +
 \Bigl( {\rm Tr}\bigl( X_n^{-1} X_{n+1/2} \bigr) \Bigr)^2
-{\rm Tr}\Bigl( \bigl( X_n^{-1} X_{n+1/2} \bigr)^2 \Bigr) \Bigr]
\label{meanF}
\end{equation}

The above formula is one of the main results of this paper.
For $N=2$ one needs to work
with matrices $X_n$ of size $2$.
Computing the necessary traces one shows
that (\ref{meanF}) simplifies to formula  (\ref{meanKN}),
while (\ref{sqrtF})
allows to write down an explicit formula 
\begin{equation}
\langle \sqrt{F}\rangle_{2,K} =
 \Bigl( {\Gamma(2K) \over \Gamma(2K+\frac{1}{2})}\Bigl)^2
 \Bigl[ \frac{3}{2}
  \Bigl( {\Gamma(K+\frac{1}{2})
\over \Gamma(K)}\Bigr)^2
+\frac{1}{2} \Bigl( {\Gamma(K-\frac{1}{2}) \over \Gamma(K-1)}\Bigr)^2 \Bigr].
\label{sqrtF2}
\end{equation}
For large $K$ this average tends to unity.
This result was derived
for $K\ge N$,
and it is ill defined, e.g for $K=1$.
Interestingly, by an analytical continuation
 $K\to 1$
one obtains the correct result for averaging
over the space of  pure states, 
$\langle \sqrt{F}\rangle_{2,1} = 2/3$
in agreement with the trivial integral
$\int_0^1\sqrt{F} dF$ following from
the uniform distribution
(\ref{ditspureN}) with $N=2$.
For the Bures measure we have
$\langle \sqrt{F}\rangle_{2,3/2} =2^7 13/(3^2 5^2\pi^2)\approx 0.7493$,
while the average for the HS measure ($K=2$)
is larger,
$\langle \sqrt{F}\rangle_{2,2} = 2^5 31/(5^2 7^2)\approx 0.8098$.

\begin{figure} [htbp]
   \begin{center}
 \includegraphics[width=9.0cm,angle=0]{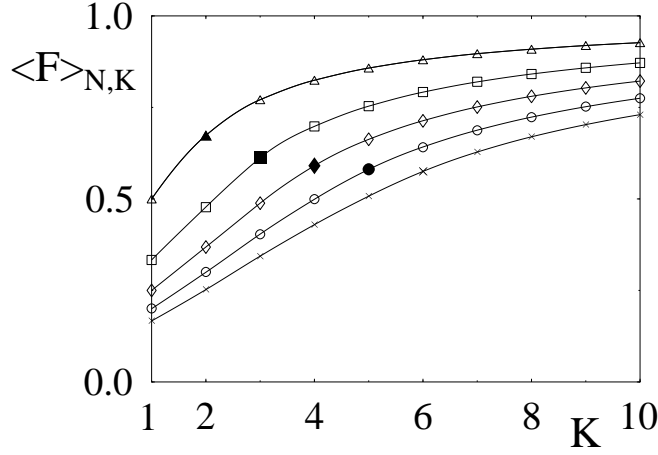}
\caption{Mean fidelity averaged over both states
distributed according to the induced measure $\mu_{N,K}$,
as a function of the size $K$ of the ancilla
for $N=2$ ($\triangle$), $N=3$ ($\square$),
$N=4$ ($\Diamond$), $N=5$ ($\circ$) and $N=6$ ($\times$).
Solid lines denote analytical results, 
 Eq. (\ref{meanKN}) and   (\ref{meanF}).
Full symbols represent averaging over 
the Hilbert-Schmidt measure, $K=N$.}
 \label{fig2}
\end{center}
 \end{figure}

Analogous results obtained for $N=3$
from  (\ref{sqrtF}) and (\ref{meanF})
 read
\begin{equation}
\langle \sqrt{F}\rangle_{3,K} =
 \Bigl( {\Gamma(3K) \over \Gamma(3K+\frac{1}{2})} \Bigr)^2
\Bigl[ \frac{3}{8}
\Bigl( {\Gamma(K-\frac{3}{2})
\over \Gamma(K-2)} \Bigr) ^2 +
\frac{3}{4} \Bigl( {\Gamma(K-\frac{1}{2}) \over
\Gamma(K-1)} \Bigr)^2
+\frac{15}{8}
\Bigl( {\Gamma(K+\frac{1}{2}) \over \Gamma(K)} \Bigr)^2 \Bigr]
\label{sqrtF3}
\end{equation}
and
\begin{equation}
\langle F\rangle_{3,K} =
\frac{1}{3} + \frac{1}{K^2} 
\Bigl[ 
\Bigl( {\Gamma(K- \frac{1}{2}) \over \Gamma(K-1)} \Bigr)^2
\Bigl[
\frac{5}{12} 
\Bigl( {\Gamma(K+\frac{1}{2}) \over \Gamma(K)} \Bigr)^2 
+\frac{1}{12}\Bigl( {\Gamma(K-\frac{3}{2})
\over \Gamma(K-2)}\Bigr)^2
\Bigr]
+\frac{1}{6}
 \Bigl( 
 \frac{\Gamma(K+\frac{1}{2}) \Gamma(K-\frac{3}{2}) }
      {\Gamma(K) \Gamma(K-2) } \Bigr)^2
\Bigr]   .
\label{meanF3}
\end{equation}
In the limit $K\to \infty$ 
both results tend to unity, e.g.
$\langle F\rangle_{3,K}\to \frac{1}{3}+
\frac{5}{12}+\frac{1}{12}+\frac{1}{6}=1$.
As in the case $N=2$,
the above formulae hold for  $K\ge N$,
but by analytical continuation
for $K\to 1$
one obtains correct results for averaging
over the space of for pure states
$\langle \sqrt{F} \rangle_{3,1} = 8/15$
and  $\langle F\rangle_{3,1} =1/3$.
Also results in the limit  $K \to 2$,
$\langle F\rangle_{3,2}= 1/3 +15(\pi/32)^2\approx 0.4779$
agree with the data obtained
numerically by averaging over the manifold of
the mixed states with rank $K=2$.
Numerical results obtained from a sample of $10^6$
random states, presented in Fig. \ref{fig2},
confer with predictions 
obtained by means of the general formula (\ref{meanF}). Interestingly,
the mean fidelity for the HS measure, 
$K=N$, decreases with the system size $N$.

\subsection {General case, $N \ge 3$, probability distributions}

Figure 3 presents the distributions of fidelity $P_{N,K}(F)$,
averaged symmetrically over both states distributed according
to the induced measures $\mu_{N,K}$ for $N=2,3$ and $4$.
Analytical results (\ref{ditspureN})
available for pure states ($K=1$)
are marked by solid lines. In general, the larger size $K$
of the ancilla, the more the distribution is shifted toward
higher values of $F$, since both states get closer to the
maximally mixed state. For large size of the ancilla 
the distributions become concentrated
at the mean value $\langle F\rangle_{N,K}$
which tends to unity in the limit $K\to \infty$.

\hskip -2.0cm
\begin{figure} [htbp]
\hskip -2.0cm
   \begin{center}
 \vskip -0.4cm
\hskip -2.0 cm
 \includegraphics[width=16.0cm,angle=0]{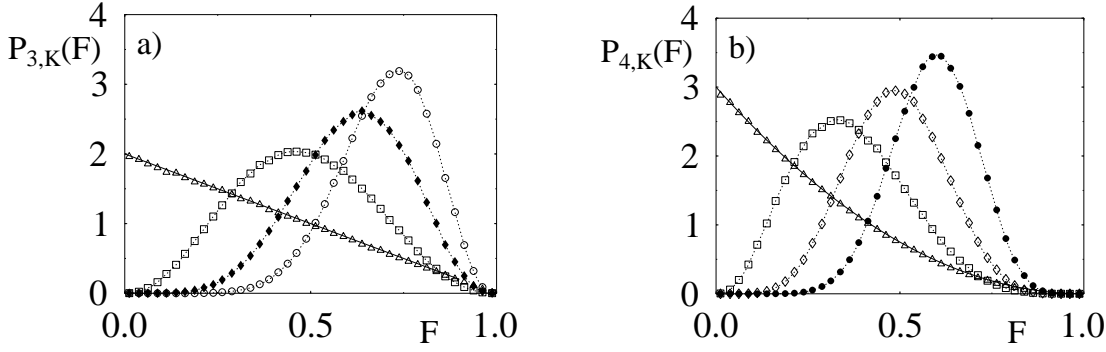}
 \vskip -0.2cm
\caption{Probability distribution of fidelity
 while averaging symmetrically over both states
distributed according to induced measures $\mu_{N,K}$.
The size of the system equals a) $N=3$ and b) $N=4$ and
while the size $K$ of the ancilla
is coded by symbols:
$K=1$ ($\triangle$), $K=2$ ($\square$),
$K=3$ ($\Diamond$) and $K=4$ ($\circ$).
Full symbols represent averaging over Hilbert--Schmidt measure, $K=N$.
Solid line for $K=1$ denote Eq. (\ref{ditspureN}),
dotted lines are plotted to guide the eye.}
 \label{fig3}
\end{center}
 \end{figure}

Obtaining explicitly analytical results
for the probability distribution $P_{N,K}(F)$
in the general case of arbitrary $N$ and $K$ 
seems not to be simple. However, we may 
obtain required information concerning these distributions
by studying the sequence of higher momenta
of the root fidelity.
Knowing all moments $\langle (\sqrt{F})^m\rangle_{N,K}$
for $m=2,4,...$ we may in principle extract the desired distribution
$P_{N,K}(F)$.
To compute such moments for $K\ge N$ we 
construct a generating function 
\begin{equation}
Z(\lambda) = \det \Bigl( \sum_{m=0}^{\infty}
\frac{(-\lambda)^m}{m!}
\Gamma\bigl(\frac{m}{2}+k+l+K-N-1\bigr)
\Gamma\bigl(\frac{m}{2}+l+K-N\bigr)
\Bigr)
\label{generN}
\end{equation}
which we obtain by replacing the fixed trace ensembles for $\rho_1$ and
$\rho_2$ by Laguerre ensembles and integrating out eigenvectors of
$\sqrt{\rho_1} \rho_2 \sqrt{\rho_1}$ with the help of the Itzykson--Zuber
integral.
The determinant concerns a $N \times N$ matrix
with indices, $k,l=1,..,N$. 
To obtain the mean values of $(\sqrt{F})^m$
one needs to multiply the expansion coefficients of $Z(\lambda)$  by 
 constants $G(m)$ defined in  (\ref{constG}):
\begin{equation}
\langle (\sqrt{F})^m\rangle_{N,K}=G(m)\ (-d/d\lambda)^m\
Z(\lambda)/Z(0)|_{\lambda =0}
\label{moments}
\end{equation}

For any fixed $N$ the moments are analytic in $K$,
so one may try to use the analytic extension for the cases
$K<N$. Alternatively, we may treat this case separately,
investigating another generating function
\begin{equation}
Z_K(\lambda) = \det \Bigl( \sum_{m=0}^{\infty}
\frac{(-\lambda)^m}{m!}
\bigl[\Gamma\bigl(\frac{m}{2}+k+N-K\bigr)\bigr]^2
\  \frac{\Gamma\bigl(\frac{m}{2}+k+l-1\bigr)}
     {\Gamma\bigl(\frac{m}{2}+k+l+N-2K\bigr)}
\Bigr)
\label{generK}
\end{equation}
with determinant of a $K \times K$ matrix
with indices, $k,l=1,..,K$.
This formula holds for $N\ge 2K$,
but may be extended analytically in $N$ beyond this
 restriction. It is obtained by reducing 
$\sqrt{\rho_1} \rho_2 \sqrt{\rho_1}$ by unitary rotations to the spaces
 of nonzero eigenvalues with dimension $K\le N$ using the symmetry properties
 of the fidelity and again integrating out all unitary rotations 
with the help of the Itzykson--Zuber integral, (here applied two times).

To show an application of this approach 
let us expand the generating function $Z_K$
up to the second order in $\lambda$. Fixing $K=2$
we obtain the averages for variable $N$
\begin{equation}
\langle \sqrt{F}\rangle_{N,2} =
\frac{\sqrt{\pi}}{16}\ (22N-13)\
 \Bigl( \frac{\Gamma(2N)} {\Gamma\bigr(2N+\frac{1}{2}\bigl)} \Bigr)^2
  \ \frac{\Gamma\bigl(N-\frac{1}{2}\bigr)} {\Gamma (N)} 
\label{sqrtFK2}
\end{equation}
and 
\begin{equation}
\langle F\rangle_{N,2} =
\frac{1}{N} \Bigl[ 1+ \frac{3 \pi }{16}\
 \frac{\Gamma \bigr(N-\frac{1}{2}\bigl)\Gamma \bigr(N+\frac{1}{2}\bigl)}
      {\Gamma (N-1) \Gamma (N+1)} \Bigr]
\label{FK2}
\end{equation}
These formulae are valid for arbitrary $N$, and for 
$N=2,3$ are consistent with Eqs. (\ref{sqrtF2})- (\ref{meanF3}).
In a similar way one may proceed with $K=3,4...$.
Furthermore, expanding the generating function to the $m$-th order
one could calculate the expectation values of $(\sqrt{F})^m$
and obtain further information concerning the 
probability distribution $P_{N,K}(F)$. There are analogous formulas for
 the moments $\langle (\sqrt{F})^m\rangle_{N,K}$ to Eqs. (\ref{sqrtF}),
(\ref{meanF}) 
in terms of traces involving $K\times K$ matrices.

To give an explicit expression for the distribution of fidelity 
$P_{N,K}(F)=W(\sqrt{F})/2\sqrt{F}$ 
we start from the fixed trace ensemble with
traces $ t_1= {\rm Tr} \rho_1$, $ t_2= {\rm Tr} \rho_2$, $ t_3= {\rm Tr}
(\sqrt{\rho_1} \rho_2 \sqrt{\rho_1})^{1/2} $. We take the Laplace transform
with respect to $t_1,t_2,t_3$, calculate it and then apply the inverse
Laplace transform at $t_1=1$, $t_2=1$, and $t_3=\sqrt{F}$. What we obtain is a
threefold contour integral
\begin{equation}
W(\sqrt{F})=C \int {ds_1 {\rm e}^{s_1} \over 2\pi i {s_1}^{KN}}\ \int {ds_2
{\rm e}^{s_2} \over 2\pi i {s_2}^{KN}}\ \int {ds_3 {\rm e}^{\sqrt{F} s_3}
\over 2\pi i  }\ Z({s_3 \over \sqrt {s_1 s_2}})\ .
\label{WvonF}
\end{equation}
with $C=\Gamma(K^2N^2)/Z(0)$. It is easy to see, that this equation is
equivalent to Eq. (\ref{moments}).
All three contours go along the whole imaginary axis with a small positive
real part.
We need the analytic properties of the function $Z(x)$ which can be seen
from the form
\begin{equation}
Z(x)=\det \left(2\int_0^{\infty} dy {y^{k-2}\Gamma[k-1+2(K-N+l)] \over
(y+1/y+x)^{k-1+2(K-N+l)}}\right)\ .
\label{Zvonx}
\end{equation}
$Z(x)$ has a cut along the real axis for $x\le -2$ and behaves
asymptotically for $x \to \infty$ like $(\ln x)^N /x^{2KN}$, which is not
easy to see, but can be derived from an explicit form for $Z(x)$
\begin{equation}
 Z(x)= \det \bigl[  (d/dx)^{2(K-N+l-1)+k} H_k(x) \ln (x/2+\sqrt{x^2/4-1})
\bigr]
\label{ZvonxEXP}
\end{equation}
with
\begin{equation}
  H_k(x)=-{\Bigl[ \bigl(x/2+\sqrt{x^2/4-1}\bigr)^{k-1} +
          \bigl(x/2-\sqrt{x^2/4-1}\bigr)^{k-1}\Bigr] } / \sqrt{x^2/4-1}
\label{H(x)}
\end{equation}
For later use we need also for $x>2$
(in the following $ \epsilon > 0$)
\begin{equation}
\lim_{\epsilon \to 0} \bigl[ Z(-x-i\epsilon)\bigr]=
      \det\bigl[ (d/dx)^{2(K-N+l-1)+k} H_k(x) \ln
\bigl(x/2+\sqrt{x^2/4-1}\bigr)
           +i \pi (d/dx)^{2(K-N+l-1)+k} H_k(x) \bigr]\ .
\label{Zvonx-i}
\end{equation}
Using appropriate contour deformations and partial integrations we are able
to perform the $s_1$ and $s_2$ integrations arriving at:
\begin{equation}
W(\sqrt{F}) =  \frac{\Gamma(K^2N^2)}{Z(0)\pi^2} \lim_{\epsilon \to 0} \Bigl[
\int_{2 }^{\infty} dx B(x )   {\rm Im}
[Z(-x-i \epsilon)]  \Bigr]
\label{WvonF1}
\end{equation}
 with
\begin{equation}
B(x)=F^{KN-1}\int_{2}^{2/\sqrt{F}} {dy \over \sqrt{1-Fy^2/4}}{\theta(x-y)\
(x-y)^{2KN-3} \over (2KN-3)!}  \ .
\label{Bvonx}
\end{equation}
Finally for numerical convenience we may use the following integral
representations for $x>2$: $Z(x)=\det(A_1)$ and 
$\lim_{\epsilon\to 0}[Z(-x-i\epsilon)]
=\det(A_1-iA_2)$ with
\begin{equation}
 A_1=4\int_{0}^{\infty} {du \cosh[(k-1)u] \Gamma[k-1+2(K-N+l)] \over
[2\cosh(u)+x]^{k-1+2(K-N+l)}}
\label{A_1}
\end{equation}
and
\begin{equation}
A_2=4\int_{0}^{\pi} {du \cos[(k-1)u] \Gamma[k-1+2(K-N+l)] \over
[2\cos(u)+x]^{k-1+2(K-N+l)}},
\label{A_2}
\end{equation}
where $k,l=1,\dots,N$, so $A_1$ and $A_2$ are square matrices
of size $N$.

\section{Concluding remarks}

In this work we have posed and solved the problem of computing
the average fidelity between two random quantum states.
As easy to predict, the results depend heavily on the probability measure,
according to which the quantum states are generated.
We have concentrated on two cases, we consider to be the most
important: the Bures measure $\mu_{\rm B}$, used to guess a random state 
of size $N$ in lack of any other information,
and the induced measure $\mu_{N,K}$, applied 
if it is known that the mixed state is obtained by partial
trace over a $K$ dimensional environment. The latter case
reduces to the Hilbert-Schmidt measure if $K=N$.

In general, one has to distinguish the case
of the symmetric averaging, in which both states are generated according to the
same measure, and the asymmetric case, e.g. the problem
of computing the mean fidelity between a pure random state and a mixed random
state.

In particular we have obtained explicit results for symmetric averaging over
the induced measures $\mu_{N,K}$, which have a simple physical
interpretation. 
The mean value $\langle F\rangle_{N,K}$
corresponds to the case of picking at random pure states  in
$N\times K$ dimensional Hilbert space and then 
studying the fidelity between both mixed states obtained by 
partial tracing over the $K$--dimensional environment.

In certain cases we have derived 
the probability distribution for fidelity $P(F)$
which may be used to formulate statistical 
statements which evaluate quantitatively 
certain quantum operations and quantum protocols.
For instance, the universal protocol of quantum cloning
of a $N$ dimensional pure state gives the fidelity
$F_u=(N+3)/(2N+2)$ \cite{We98}.
Making use of the distribution (\ref{ditspureN})
we see that the probability $p_N$ that
a randomly taken $N$ dimensional pure state  will give a better fidelity
than this obtained by the cloning procedure 
is equal 
\begin{equation}
p_N=\int_{F_u}^1  P_{N,1}(F)dF =
\frac{1}{2^{N-1}} \
\Bigl(\frac{N-1}{N+1} \Bigr)^{N-1} .
\label{pfidrand}
\end{equation}
For $N=2$ there exists a considerable probability,
$p_2=1/6$, that picking at random
a pure state of a qubit provides results better 
then the cloning protocol. However, for larger 
system sizes the probability $p_N$
decreases exponentially with $N$, which means
that the relative strength of the cloning procedure with respect to
the 'random choice' strategy, increases with the size of the system.

We thank D. Savin for useful discussions
and I. Bengtsson, R. Mu{\`n}oz-Tapia and A. Uhlmann 
for helpful correspondence.
This work was supported by Sonderforschungsbereich / Transregio 12
der Deutschen Forschungsgemeinschaft and a solicitated grant
number PBZ-MIN-008/P03/2003
of Polish Committee for Scientific Research.

\appendix
\section{Generating functions for moments}

\subsection{ Induced measures: Wishart case, $K \ge N $}

In this appendix we present a brief derivation of the
generating functions for moments of the square-root fidelity
which hold for an arbitrary induced measure $\mu_{N,K}$. In the 
first subsection we consider the case $K \ge N $.
 Then the distribution of square-root fidelity is given by
\begin{equation}
W(\sqrt{F}) \propto \int D\rho_1\ \int D\rho_2\ 
\det(\rho_1 \rho_2)^{K-N} \delta(1-{\rm Tr}\rho_1)\ 
\delta(1-{\rm Tr}\rho_2)\ 
\delta \Bigl( \sqrt{F}-{\rm
Tr}\bigl(\sqrt{\rho_1}\rho_2\sqrt{\rho_1}\bigr)^{1/2}\Bigr) ,
\label{WK>=N}
\end{equation}
where $D\rho_1$, $D\rho_2$ are the matrix volume
 elements and we have to integrate over positive matrices
 $\rho_1$ and $\rho_2$. Since ${\rm
Tr}((\sqrt{\rho_1}\rho_2\sqrt{\rho_1})^{1/2})$ 
is a homogeneous function of both, $\rho_1$ and $\rho_2$ of degree $1/2$,  
moments can equivalently be obtained from the generating function
\begin{equation}
Z(\lambda ) \propto \int D\rho_1\ \int D\rho_2\ \det(\rho_1 \rho_2)^{K-N} 
{\rm e}^{-{\rm Tr}\rho_1}\ {\rm e}^{-{\rm Tr}\rho_2}\ 
{\rm e}^{-\lambda {\rm Tr}[(\sqrt{\rho_1}\rho_2\sqrt{\rho_1})^{1/2}]}
\label{Zvonl}
\end{equation}
following Equ. (\ref{moments}). We make a transformation 
$\rho_2 $   $\to $  $ {\rho_1}^{-1/2}\rho_2{\rho_1}^{-1/2}$ and taking 
into account the Jacobian we obtain
\begin{equation}
Z(\lambda ) \propto \int D\rho_1\ \int D\rho_2\ \det(\rho_1)^{-N} 
\det(\rho_2)^{K-N} {\rm e}^{-{\rm Tr}\rho_1}\ {\rm e}^{-{\rm Tr}\rho_2/\rho_1}\  
{\rm e}^{-\lambda {\rm Tr}(\rho_2)^{1/2}}\ .
\label{Zvonl1}
\end{equation}
Note that the nasty square roots disappeared and we can use the 
Itzykson--Zuber integral to integrate out the eigenvectors of $\rho_2$ in 
the expression ${\rm e}^{-{\rm Tr}\rho_2/\rho_1}$. As result we obtain
\begin{equation}
Z(\lambda ) \propto \int_0^{\infty} {dx_1 \over x_1}...{dx_N \over x_N}
 \int dy_1...dy_N (y_1...y_N)^{K-N}   
  {\rm e}^{-(x_1+...x_N)}\ {\rm e}^{-\lambda (\sqrt{y_1}+...\sqrt{y_N})}\ 
\det({\rm e}^{-y_k/x_l})    
 \prod_{k<l}(x_k-x_l)(y_k-y_l)\  
\label{Zvonl2}
\end{equation}
where $x_1,...x_N$ are the eigenvalues of $\rho_1$ and $y_1,...y_N$ 
are the eigenvalues of $\rho_2$. Using the properties of the Vandermonde 
determinant we reduce this expression to a single determinant
 and arrive at Eq. (\ref{Zvonx}). 

\subsection{Induced measures: anti--Wishart case, $K <N $}
In this case the matrices $\rho_1$ and $\rho_2$ are of rank $K$ and they can be
brought by a unitary transformation of the form
 $$  U= \left(
\begin{matrix} \sqrt{1-XX^{\dagger}}        & -X                   \cr
                        X^{\dagger}         & \sqrt{1-X^{\dagger}X}\cr
 \end{matrix}  \right)     $$
to block diagonal form. If $U$ is drawn from CUE it can be shown that the
corresponding 
measure $DX$ is flat, where $X$ is a complex $K\times(N-K)$ matrix with the 
restriction $XX^{\dagger}\le 1$. Calling the nonzero blocks again $\rho_1$ and
$\rho_2$, 
now $K\times K$ matrices, we may write the generating function for moments as
\begin{equation}
Z_K(\lambda ) \propto \int D\rho_1\ \int D\rho_2\ 
\int_{XX^{\dagger}\le 1}DX\ \det(\rho_1 \rho_2)^{N-K} {\rm e}^{-{\rm
Tr}\rho_1}\ 
{\rm e}^{-{\rm Tr}\rho_2}\  {\rm e}^{-\lambda {\rm Tr}[(\sqrt{\rho_1}
\sqrt{1-XX^{\dagger}}\rho_2\sqrt{1-XX^{\dagger}}\sqrt{\rho_1})^{1/2}]}
\label{Zvonl3}\ .
\end{equation}
In a first step we make the transformation $\rho_2 $   $\to $ 
 $ (1-XX^{\dagger})^{-1/2}\rho_2(1-XX^{\dagger})^{-1/2}$ arriving at
\begin{equation}
Z_K(\lambda ) \propto \int D\rho_1\ \int D\rho_2\ \int_{XX^{\dagger}\le 1}DX\
{\det(\rho_1 \rho_2)^{N-K} \over \det(1-XX^{\dagger})^N} {\rm e}^{-{\rm
Tr}\rho_1}\ {\rm e}^{-{\rm Tr}\rho_2/(1-XX^{\dagger})}\ 
 {\rm e}^{-\lambda {\rm Tr}[(\sqrt{\rho_1} \rho_2 \sqrt{\rho_1})^{1/2}]}
\label{Zvonl4}\ .
\end{equation}
Then we use the symmetry ${\rm Tr}((\sqrt{\rho_1} \rho_2
\sqrt{\rho_1})^{1/2})={\rm Tr}((\sqrt{\rho_2} \rho_1 \sqrt{\rho_2})^{1/2})$
and make the transformation $\rho_1 $   $\to $  $
{\rho_2}^{-1/2}\rho_1{\rho_2}^{-1/2}$ leading to
\begin{equation}
Z_K(\lambda ) \propto \int D\rho_1\ \int D\rho_2\ \int_{XX^{\dagger}\le 1}DX\
{\det(\rho_1  )^{N-K} \over \det(\rho_2)^K \det(1-XX^{\dagger})^N} {\rm
e}^{-{\rm Tr}\rho_1/\rho_2}\ {\rm e}^{-{\rm Tr}\rho_2/(1-XX^{\dagger})}\  {\rm
e}^{-\lambda {\rm Tr}\sqrt {\rho_1}}  
\label{Zvonl5}\ .
\end{equation}
Now we see that the matrix $W=X X^{\dagger}$ is again of the Wishart type
and is positive definite for $K\le(N-K)$. Let us focus on this case, then the
distribution of $W$ is $\propto \det(W^{N-2K})$ and we may reduce the
integral to the eigenvalues $x_1,...x_K$ of $\rho_1$, $y_1,...y_K$ of $\rho_2$
and ${z_1,...z_K}$ of $W$ using two times the Itzykson--Zuber integral. The
resulting expression contains in the integrand the Vandermonde determinants
of
$x_1,...x_K$ and of $z_1,...z_K$ and can therefore again be reduced to a
single determinant which is given by
\begin{equation}
Z_K(\lambda )= \det\left( 2\int_0^{\infty}dx\int_0^{\infty}{dy\over y}
\int_0^1dz{x^{N-K+k-1}z^{N-2K} \over (1-z)^{N-K-l+1}} 
\exp\left[-\sqrt{{x\over 1-z}}\left(y+{1\over y}\right)\right]
 {\rm e}^{-\lambda\sqrt{x}}\right)    
\label{Zvonl6}\ .
\end{equation}
Expanding the threefold integral in powers of $\lambda$, one arrives at
Eq. (\ref{generK}).

\section{Derivation of the distribution (\ref{ditsmixBuN})}

To derive the distribution $P(F)$ for the fidelity between 
a random pure state and a random mixed state
distributed according to the Bures measure (\ref{mesbur2})
consider 
\begin{equation}
P({F}) \propto \int_{\rho>0} D\rho\ \int_{A=A^{\dagger}}DA \  
 \delta(1-{\rm Tr}\rho A^2)\ 
\delta(1-{\rm Tr}\rho)\ 
\delta  (  {F}-{\rm Tr}\rho |\psi \rangle \langle \psi | ),
\label{PB}
\end{equation}
where we introduced an auxiliary Hermitian matrix $A$ to generate the distribution 
(\ref{mesbur2}) and the last two $\delta$-functions correspond to the constraints 
(\ref{trace}) and (\ref{fidpure}). The $\delta$-functions may be written as inverse 
Laplace transforms along the slightly to the right shifted imaginary axis and 
then the integration over the positive definite $N\times N$-matrix $\rho$ can be done arriving at
\begin{equation}
P({F})\propto  \int {ds_1 {\rm e}^{s_1} \over 2\pi i  }\ \int {ds_2
{\rm e}^{s_2} \over 2\pi i}  \ \int {ds_3 {\rm e}^{{F} s_3}
\over 2\pi i  }\ \int DA \det (s_1A^2 + s_2 +s_3 |\psi \rangle \langle \psi |)^{-N}\ .
\label{PvonF}
\end{equation}
After a suitable rescaling of the matrix A, we can do the $s_1$- and $s_3$-integrations and obtain
\begin{equation}
P({F})\propto \int{ DA \over \det(1+A^2)^N} 
\int_{-i\infty + \epsilon}^{+i\infty + \epsilon} {ds {\rm e}^{s} \over 2\pi i s^{N^2/2-N} }\ 
(\langle \psi | (1+A^2)^{-1} |\psi \rangle)^{-N}\ F^{N-1}\ 
\exp\Bigl( -{s F \over \langle \psi | (1+A^2)^{-1} |\psi \rangle}\Bigr)  \ .
\label{PvonF2}
\end{equation}
Now it is possible to show, that with the correspondence $U=(1+iA)/(1-iA)$
the measure $DA/\det(1+A^2)^N$ is equivalent to the invariant measure on the   group of unitary $N\times N$-matrices $U$. Since only $\langle \psi |(1+A^2)^{-1} |\psi \rangle=\langle \psi | (2+U+U^{\dagger})/4 |\psi \rangle$ occurs, the integration over $A$ can be reduced to a one-dimensional one and after integrating over $s$ in the complex plane and restoring the normalization we end up with formula (\ref{ditsmixBuN}).

\end{document}